\documentclass[pra,aps,twocolumn,superscriptaddress,showpacs]{revtex4}
\usepackage[dvips]{graphicx}

\begin{document}
\author{Wen-Long Yang}
\affiliation{Theoretical Physics Division, Chern Institute of
Mathematics, Nankai University, Tianjin 300071, P.R.China}
\author{Jing-Ling Chen}
\email{chenjl@nankai.edu.cn} \affiliation{Theoretical Physics
Division, Chern Institute of Mathematics, Nankai University, Tianjin
300071, P.R.China}
\title{Relation between three-qubit entanglement invariants and two-qubit concurrence}
\begin{abstract}
In this Brief Report we show the relation between three-qubit
entanglement invariants and two-qubit concurrence with the help of
projective measurements. How to use these invariants to represent
the entanglement property of three-qubit pure states is also
investigated.
\end{abstract}
 \maketitle

Quantum entanglement and quantum nonlocality are regarded as a key
feature of quantum mechanics since its early days \cite{35EPR,64Be}.
Recently, by taking advantage of the entanglement property there has
emerged an attractive research field called quantum information,
which is composed of quantum cryptography, quantum teleportation,
quantum computation and so on \cite{00NiCh}. Frequent use of
entanglement has imminently called for a precise and proper
description of entanglement for quantum states. For two-qubit state,
entanglement entropy and entanglement of formation and the
well-known concurrence
\cite{96BBPS,96BDSW,97PoRo,97VPRK,97HiWo,98Wo} are all good measures
for the degree of entanglement for two-qubit states. During the
period of finding a quantity as a measure of the degree of
entanglement for three-qubit states, five
local-unitary-transformation invariants have been thought to be
important resources to form such a quantity
\cite{96ScMa,98LiPo,98GRB,99Ke,00Su,05Le}. Efforts have also been
devoted to find such a quantity to measure the entanglement of
three-qubit pure states from the viewpoint Bell inequality
\cite{04CWKO}. However, three qubits can be entangled in two
inequivalent ways, namely the Greenberger-Horne-Zeilinger (GHZ)
state \cite{89GHZ} and the $W$ state \cite{00DVC}, therefore finding
only one unified quantity to describe the entanglement of all
three-qubit states seems to be not easy.

In 2002, Linden \emph{et al.} pointed out that almost every pure
state of three qubits is completely determined by its two-particle
reduced density matrices \cite{02LPW}. This may hint that there is a
definite relation between five local-unitary-transformation
entanglement invariants of three-qubit between two-qubit
concurrence. To our knowledge, such a relation has not been reported
in the literature before. The purpose of this Brief Report is to
show the exact relation between the three-qubit entanglement
invariants and the two-qubit concurrence with the help of the
projective measurements.

The report is organized as follows. First, we show the relation
between local-unitary-transformation invariants of three-qubit pure
states and concurrence of two-qubit pure states via projective
measurements. Then, we investigate how to use these invariants to
represent the entanglement property of three-qubit pure states.
Conclusion is made in the end.

A two-qubit state is the spin state or pseudo-spin state composed of
two spin-$1/2$ particles $A$ and $B$, which reads
\begin{eqnarray}
\label{twoqubit}\left|\psi_2\right>=\mu_{00}\left|00\right|+
\mu_{01}\left| 01\right>+\mu_{10}\left|10\right>+\mu_{11}\left|
11\right>.
\end{eqnarray}
The state is separable means that it can be written in the following
direct-product form
\begin{eqnarray}
\label{separable}\left(a\left|0\right>_A+b\left|1\right>_A\right)\otimes\left(c\left|0\right>_B+d\left|
1\right>_B\right),
\end{eqnarray}
otherwise it is entangled or inseparable. Concurrence is one of
proper measures of the degree of entanglement for two-qubit states.
It is first defined in \cite{97HiWo} to find a simple way to
quantify the entanglement of two-qubit mixed state $\rho_{AB}$
\begin{eqnarray}
C=max\{\lambda_1-\lambda_2-\lambda_3-\lambda_4,0\},
\end{eqnarray}
where $\lambda_i$ is the square root of eigenvalues of
$\rho_{AB}\cdot(\sigma_y\otimes\sigma_y)\cdot\rho_{AB}^*\cdot
(\sigma_y\otimes\sigma_y)$ in decreasing order. The definition is
also work for pure-state case of two qubits, in this case it
possesses a simpler form
\begin{eqnarray}
\label{pureC2}C=2\sqrt{\det(\rho_A)}=2\sqrt{\det(\rho_B)},
\end{eqnarray}
where $\rho_A$ and $\rho_B$ are density matrices obtained from the
pure state $\rho_{AB}$ by tracing out the other particle. The
concurrence (\ref{pureC2}) is a local-unitary-transformation
invariant for $\rho_{AB}$ which means $C(\rho_{AB})=C((U_A\otimes
U_B) \rho_{AB}(U_A\otimes U_B)^\dag)$. For two-qubit pure state
(\ref{twoqubit}), the concurrence is usually expressed in terms of
the coefficients $\mu_{ij}$ as
\begin{eqnarray}
\label{Con}C=2\left| \mu_{00} \mu_{11} - \mu_{01} \mu_{10} \right|.
\end{eqnarray}
$C$ is $0$ for separable state (\ref{separable}) and $1$ for
maximally entangled state
$(1/\sqrt{2})(\left|00\right>+\left|11\right>)$.

Local-unitary-transformation invariants for three-qubit states have
been found for about one decade
\cite{96ScMa,98LiPo,98GRB,99Ke,00Su}. For a three-qubit pure state
\begin{eqnarray}
\label{3qs}\left| \psi
\right> = \sum \limits_{i,j,k=0,1} \mu_{ijk} \left| ijk \right>,
\end{eqnarray}
the indices $i$, $j$ and $k$ stand for the spin states of three
spin-$1/2$ particles $A$, $B$ and $C$. More precisely, the five
linear independent entanglement invariants (i.e., $I_1$ to $I_5$)
are \cite{00Su,00CKW}
\begin{eqnarray}
\label{I}I_0&=&\left< \psi \right| \left. \psi \right>,\nonumber \\
I_1&=&{\rm Tr}(\rho_A^2),\nonumber \\
I_2&=&{\rm Tr}(\rho_B^2),\nonumber \\
I_3&=&{\rm Tr}(\rho_C^2),\nonumber \\
I_4&=&3{\rm Tr}(\rho_A\otimes \rho_B . \rho_{AB})- {\rm
Tr}(\rho_{A}^3) -
{\rm Tr}(\rho_{B}^3),\nonumber \\
I_5&=&16\left|\mu_{000}^2\mu_{111}^2+\mu_{001}^2\mu_{110}^2 +
\mu_{010}^2 \mu_{101}^2 + \mu_{011}^2 \mu_{100}^2\right.\nonumber \\
&-& 2(\mu_{000} \mu_{111} \mu_{011} \mu_{100} + \mu_{000} \mu_{111}
\mu_{101} \mu_{010} \nonumber \\
&+& \mu_{000} \mu_{111} \mu_{110}\mu_{001} + \mu_{011}
\mu_{100}\mu_{101} \mu_{010}\nonumber \\
&+& \mu_{011} \mu_{100} \mu_{110} \mu_{001} + \mu{101} \mu_{010}
\mu_{110}
\mu_{001})\nonumber \\
&+& \left.4(\mu_{000} \mu_{011} \mu_{101} \mu_{110} + \mu_{111}
\mu_{001} \mu_{010} \mu_{100}) \right|^2,
\end{eqnarray}
where $I_0$ is the identity and a homogeneous polynomial of degree
two. $I_1$, $I_2$ and $I_3$ are homogeneous polynomials of degree
four, $I_4$ is a homogeneous polynomial of degree six and $I_5$ is a
homogeneous polynomial of degree eight.

Though there are several generalizations of two-qubit concurrence to
multi-particle concurrence \cite{05MKB,06He}, the relation between
 (\ref{Con}) and (\ref{I}) has not yet been revealed before.  In the following
we would like to show an intimate relation between them.

We first make a projective measurement $|\psi_A \rangle \langle
\psi_A|$ with respect to particle $A$ with
\begin{eqnarray}
\label{statea}\left| \psi_A \right> &=& \cos \frac{\theta_A}{2}
\left| 0 \right>_A + \sin \frac{\theta_A}{2} e^{i \phi_A} \left| 1
\right>_A.
\end{eqnarray}
This is like measuring the spin of $A$ in the direction $(\sin
\theta_A \cos \phi_A,\sin\theta_A\sin\phi_A,\cos\theta_A)$. After
performing the measurement, one gets the collapsed state as
\begin{eqnarray}
\label{measuredstate}&&\left| \psi_{BC} \right>= \frac{\left<\psi_A
\right.\left|\psi \right>}{\sqrt{W_{BC}(\theta_A,\phi_A)}},
\end{eqnarray}
where
\begin{eqnarray}
\label{W}W_{BC}(\theta_A,\phi_A)=\left< \psi \right| \left.\psi_A
\right> \left< \psi_A  \right.\left| \psi \right>.
\end{eqnarray}
Namely, after this measurement, one gets a pure two-qubit state
$\left| \psi_{BC} \right>$  with the probability
$W_{BC}(\theta_A,\phi_A)$. We can calculate the concurrences
$C_{BC}(\theta_A,\phi_A)$ of $\left| \psi_{BC}\right>$ with the
definition (\ref{Con}). $C_{BC}(\theta_A,\phi_A)$ is an invariant
under local unitary transformation of $B$ and $C$, which is the
property of the concurrence. From (\ref{W}), one may show that
$W_{BC}(\theta_A ,\phi_A)$ is also an invariant under local unitary
transformation of $B$ and $C$. The local unitary transformation of
$A$ rotates $\left|\psi_A\right>$ in its spin space, we found the
integration of $W_{BC}^2(\theta_A,\phi_A) C_{BC}^2(\theta_A,\phi_A)$
in the whole spin space of $A$ is invariant under local unitary
transformations of $A$, $B$ and $C$. Now we denote this quantity as
\begin{eqnarray}
\label{C4}C_{BC4}&=&\int^{\pi}_0 d \theta_A \int^{2\pi}_0 d \phi_A
\sin \theta_A \nonumber \\
&&W_{BC}^2(\theta_A,\phi_A) C_{BC}^2(\theta_A,\phi_A),
\end{eqnarray}
where the squared concurrence $C_{BC}^2(\theta_A,\phi_A)$ has been
adopted, such that the integrand does not contain the square-root
calculation and the integration becomes easier to perform.
Similarly, we can get $C_{AB4}$ and $C_{AC4}$. Here we have added a
$4$ in the subscript to denote that $C_{AB4}$, $C_{AC4}$ and
$C_{BC4}$ are all homogeneous polynomial invariants of degree four.
Such an approach can be generalized to obtain polynomials of higher
degree just by raising the order of the weight
$W_{BC}(\theta_A,\phi_A)$. For instances, the polynomial invariants
of degree six can be obtained through
\begin{eqnarray}
\label{C6}C_{BC6}&=&\int^{\pi}_0 d \theta_A \int^{2\pi}_0 d \phi_A
\sin \theta_A \nonumber \\
&&W_{BC}^3(\theta_A,\phi_A) C_{BC}^2(\theta_A,\phi_A),
\end{eqnarray}
and the polynomial invariants of degree eight can be achieved by
\begin{eqnarray}
\label{C8}C_{BC8}&=&\int^{\pi}_0 d \theta_A \int^{2\pi}_0 d \phi_A
\sin \theta_A \nonumber \\
&&W_{BC}^4(\theta_A,\phi_A) C_{BC}^4(\theta_A,\phi_A).
\end{eqnarray}
$C_{AB6}$, $C_{AC6}$, $C_{AB8}$ and $C_{AC8}$ can be obtained in the
same way. They are all homogeneous polynomials which are invariant
under local unitary transformation, therefore they should be able to
be expressed by the set of polynomial invariants $I$'s mentioned in
Eq. (\ref{I}). After making some observations, we found their
relations as
\begin{eqnarray}
C_{AB4}&=&\frac{\pi}{3}(-I_1 - I_2 + I_3 + I_0^2),\nonumber \\
C_{AC4}&=&\frac{\pi}{3}(-I_1+ I_2 - I_3 + I_0^2),\nonumber \\
C_{BC4}&=&\frac{\pi}{3}(I_1 - I_2 - I_3 + I_0^2),\nonumber \\
C_{AB6}&=&\frac{\pi}{18}[2I_0^3 - 3I_0(I_1+I_2-2I_3) - 2 I_4],\nonumber \\
C_{AC6}&=&\frac{\pi}{18}[2I_0^3 - 3I_0(I_1-2I_2+I_3) - 2 I_4],\nonumber \\
C_{BC6}&=&\frac{\pi}{18}[2I_0^3 - 3I_0(-2I_1+I_2+I_3) - 2 I_4],\nonumber \\
C_{AB8}&=&\frac{\pi}{240}\left[12(-I_1-I_2+I_3+I_0^2)^2-I_5\right],\nonumber \\
C_{AC8}&=&\frac{\pi}{240}\left[12(-I_1+I_2-I_3+I_0^2)^2-I_5\right],\nonumber \\
C_{BC8}&=&\frac{\pi}{240}\left[12(I_1-I_2-I_3+I_0^2)^2-I_5\right].
\label{rec8}
\end{eqnarray}
It is not hard to obtain the $I$'s from the $C$'s, i.e.,
\begin{eqnarray}
\label{CTOI}I_1&=&I_0^2-\frac{3}{2\pi}(C_{AB4}+C_{AC4}),\nonumber \\
I_2&=&I_0^2-\frac{3}{2\pi}(C_{AB4}+C_{BC4}),\nonumber \\
I_3&=&I_0^2-\frac{3}{2\pi}(C_{AC4}+C_{BC4}),\nonumber \\
I_4&=&I_0^3-\frac{9}{4\pi}\left[4 C_{AB6}  +
I_0(-2C_{AB4}+C_{AC4}+C_{BC4})\right]\nonumber \\
&=&I_0^3-\frac{9}{4\pi}\left[4 C_{AC6}  +
I_0(C_{AB4}-2C_{AC4}+C_{BC4})\right]\nonumber \\
&=&I_0^3-\frac{9}{4\pi}\left[4 C_{BC6}  +
I_0(C_{AB4}+C_{AC4}-2C_{BC4})\right],\nonumber \\
I_5&=&\frac{12}{\pi}[\pi(-I_1-I_2+I_3+I_0^2)^2-20 C_{AB8}]\nonumber \\
&=&\frac{12}{\pi}[\pi(-I_1+I_2-I_3+I_0^2)^2-20 C_{AC8}]\nonumber \\
&=&\frac{12}{\pi}[\pi(I_1-I_2-I_3+I_0^2)^2-20 C_{BC8}].
\end{eqnarray}
This is the relation between three-qubit entanglement invariants and
two-qubit concurrence via projective measurements. It is worthy to
mention that there is another way to obtain the polynomial
invariants of degree eight:
\begin{eqnarray}
\label{C8P}C'_{BC8}&=&\int^{\pi}_0 d \theta_A \int^{2\pi}_0 d \phi_A
\sin \theta_A \nonumber \\
&&W_{BC}^4(\theta_A,\phi_A) C_{BC}^2(\theta_A,\phi_A).
\end{eqnarray}
and also $C'_{AB8}$ and $C'_{AC8}$. They can also be expressed by
the $I$'s as
\begin{eqnarray}
C'_{AB8}&=&\frac{\pi}{480}\left[I_5-64 I_4 I_0+4(-I_1^2-I_2^2+7
I_3^2)\right.\nonumber \\
&&+8(-3I_1-3I_2+17I_3)I_0^2\nonumber \\
&&\left.+8(I_1 I_2 -3I_1 I_3 -I_2 I_3)-4 I_0^4\right],\nonumber \\
C'_{AC8}&=&\frac{\pi}{480}\left[I_5-64 I_4 I_0+4(-I_1^2+7I_2^2-
I_3^2)\right.\nonumber \\
&&+8(-3I_1+17I_2-3I_3)I_0^2\nonumber \\
&&\left.+8(-3I_1 I_2 +I_1 I_3 -3 I_2 I_3)-4 I_0^4\right],\nonumber \\
C'_{BC8}&=&\frac{\pi}{480}\left[I_5-64 I_4 I_0+4(7I_1^2-I_2^2-
I_3^2)\right.\nonumber \\
&&+8(-3I_1+17I_2-3I_3)I_0^2\nonumber \\
&&\left.+8(-3I_1 I_2 +I_1 I_3 -3 I_2 I_3)-4 I_0^4\right].
\end{eqnarray}
The different is that these formula are more complicated than the
previous ones as in Eq. (\ref{rec8}).

From the definitions (\ref{C4}), (\ref{C6}) and (\ref{C8}), we can
measure these $C$'s. One notices that two-qubit concurrence for pure
state can be obtained locally from (\ref{pureC2}) by measuring the
density matrix of either party of the two-qubit state. Such a
discussion has been made in \cite{00SaHu} and a recent experiment
\cite{06WRDMB} has been carried out to realize it. Moreover the
probability, e. g. $W_{BC}(\theta_A,\phi_A )$, can also be obtained
locally. Therefore for an unknown three-qubit pure state shared by
Alice, Bob and Charlie, $C_{BC}(\theta_A,\phi_A)$ is locally
measured by Bob and $W_{BC}(\theta_A,\phi_A )$ is locally measured
by Alice. After Alice scanned all $\theta_A$ and $\phi_A$, and they
tell each other the quantity they obtained through classical
channel, then they can compute $C_{BC4}$, $C_{BC6}$ and $C_{BC8}$
with (\ref{C4}), (\ref{C6}) and (\ref{C8}). By changing their roles
in turn, they will know all the invariants of the state, or say,
they know all entanglement properties of the state. So
theoretically, we can get the entanglement property of the pure
entangled three-qubit state by local measurement and classical
communication.

\begin{figure}
  \includegraphics[width=4cm]{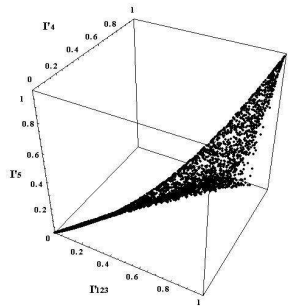}\\
\caption{$5000$ points for random three-qubit states are plotted in
the $I'_{123}-I'_4 - I'_5$ coordinate system.}   \label{I123I4I5}
\end{figure}
\begin{figure}
  \includegraphics[width=4cm]{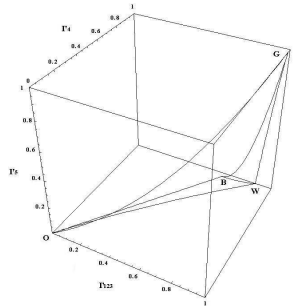}\\
\caption{This figure shows the line boundary of all pure three-qubit
states in the $I'_{123}-I'_4 - I'_5$ coordinate system. The point
$O$ stands for the states like $\left| \psi \right>_O$. The point
$B$ stands for the states like $1/\sqrt{2} \left|000\right> +
1/\sqrt{2} \left| 011 \right>$, in which one of the three particles
is not entangled with the other two maximally entangled particles.
$W$ stands for the states like $\left|\psi\right>_W$, and $G$ stands
for the states like $\left| \psi \right>_G$.}   \label{states}
\end{figure}
Now we investigate how to use these five
local-unitary-transformation invariants (\ref{I}) to represent the
entanglement properties of three-qubit pure states. For two-qubit
system, we know that the concurrence, as a
local-unitary-transformation invariant, is a good measure of degree
of entanglement. But three-qubit system has two kinds of different
entanglement, i.e., the GHZ class and the $W$ class \cite{00DVC},
thus it might be more appropriate that we use more than one quantity
to represent the entanglement property of three qubits. Luckily, we
have five local-unitary-transformation invariants. It can be seen
from (\ref{I}) that $I_1$, $I_2$ and $I_3$ are not invariant under
the permutation of the three particles and $I_4$ and $I_5$ are
invariant under such a permutation. In order to eliminate the
difference brought by the permutations between particles, we use
$I_{123}=I_1+I_2+I_3$, $I_4$ and $I_5$ as our measurements for the
entanglement of three-qubit pure states. In such a coordinate
system, when we talk about a point for a state, this point
corresponds to states which can be transformed to the state by local
unitary transformations and together with permutations of $A$, $B$
and $C$. We find that $I_{123}$ has its maximum value $3$ for the
direct-product states $\left| \psi \right>_O= \left|000 \right>$,
and its minimum value $1.5$ for the GHZ state, i.e.,
$\left|\psi\right>_G=1/\sqrt{2}(\left|000\right> + \left|
111\right>)$; $I_4$ reaches its maximum value $1$ for $\left| \psi
\right>_O$ and its minimum value $2/9$ for the $W$ state, i.e.,
$1/\sqrt{3}( \left|001\right> + \left|010\right> +\left|100
\right>$; and $I_5$ ranges from 1 for $\left|\psi\right>_G$ to $0$
for $\left|\psi\right>_W$ and $\left| \psi \right>_O$. To make the
figure looks better, we rescale three new quantities as
\begin{eqnarray}
\label{Ip}I'_{123}&=&2(3-I_{123})/3,\nonumber \\
I'_4&=&9(1-I_4)/7, \;\; I'_5=I_5,
\end{eqnarray}
which are all range from 0 to 1. An arbitrary three-qubit state has
its corresponding position in the coordinate system $I'_{123}-I'_4-
I'_5$. The origin stands for the direct-product state $\left| \psi
\right>_O$; $\left|\psi\right>_G$ has maximum 1 in the axes
$I'_{123}$ and $I'_5$, and $\left| \psi \right>_W$ has maximum 1 in
the axis $I'_4$. $I'_4$ and $I'_5$ could be the two quantities
describing two distinct three-body entanglements
\cite{96ScMa,00DVC}. To make it visible, points for $5000$ random
three-qubit states have been plotted in Fig. \ref{I123I4I5} in
$I'_{123}-I'_4 - I'_5$ coordinate system.

We also find the line boundary of Fig. \ref{I123I4I5}, which is
shown in Fig. \ref{states}. It is of interest to show the
corresponding states in the line boundary. In Fig. \ref{states}, the
line $OG$ stands for generalized $GHZ$ states $\cos \theta \left|
000 \right> + \sin \theta \left| 000 \right>$ with
$\theta\in[0,\pi/2]$. $OB$ are states $\cos \theta \left| 000
\right> + \sin \theta \left| 011 \right>$ with $\theta \in
[0,\pi/2]$. $OW$ are states like $(\cos \theta /\sqrt{3})\left(
\left| 011 \right> + \left| 101 \right> + \left| 110 \right> \right)
+ \sin \theta \left| 111 \right>$ with $\theta \in [0,\pi/2]$. $BW$
are states $(\cos \theta / \sqrt{2}) \left( \left|000\right> +
\left|111\right>\right) + (\cos \theta / \sqrt{2}) \left( \left|100
\right> + \left| 011 \right> \right)$ with $\theta \in [0,\pi]$.
$BG$ are states $(\sin \theta /\sqrt{2}) \left( \left| 101 \right> +
\left| 000 \right> \right) + \cos \theta \left| 110 \right>$ with
$\theta \in [\arctan \sqrt{2},\pi/2]$. And $WG$ are states $(\sin
\theta /\sqrt{6}) \left( \left|001\right> + \left|010\right> +
\left|100\right> + \left|011\right> + \left| 101 \right> + \left|
110 \right> \right) + (\cos \theta /\sqrt{2}) \left( \left|000
\right>+ \left|111\right>\right)$, with $\theta \in
[2/3\pi,5\pi/6]$. These state are lie in the line boundary of the
figure, it might be because they are the nearest ways (or more
precisely orbits \cite{97PoRo}) between four states $\left|
\psi\right>_O$, $\left| \psi\right>_B$, $\left| \psi\right>_W$ and
$\left| \psi\right>_G$ in the $SU(8)$ parameter space of three
qubits.

Let us see how the points in Fig. \ref{I123I4I5} represent the
entanglement properties of three-qubit states. We know that all
two-body entangled states are on $OB$. Ref. \cite{02LPW} indicated
that the generalized $W$ states
$\left|\psi\right>'_W=a\left|001\right>+b \left|010\right>+ c\left|
100\right>$ were uniquely determined by their two-party reduced
states. We can understand this from Fig. \ref{states} by calculating
their $I'_5$, and we know that all $\left|\psi\right>'_W $ have zero
$I'_5$.
Though we cannot say that the distance from a point to the origin in
the $I'_{123}-I'_4 - I'_5$ coordinate system can measure the degree
of entanglement of the three-qubit state, at least we can get a
brief impression of the entanglement property of the state. The
point on or near $OB$ will be more likely a two-particle entangled
state, and a state very near the origin should be less entangled
than those who are much far from the origin.

In conclusion, we show the relation between three-qubit entanglement
invariants and two-qubit concurrence with the help of projective
measurements. This relation also provides a method to get
three-qubit local-unitary-transformation invariants by local
measurement and classical communication.
These five invariants can form three new quantities (\ref{Ip}),
which are invariant under both local unitary transformation and
permutations. In the $I'_{123}$-$I'_4$-$I'_5$ coordinate, we can
distinguish two-body entanglement state and the generalized $W$
states from other states. For a three-qubit state, it is helpful for
us to know some of its entanglement property by plotting its
corresponding point in the $I'_{123}$-$I'_4$-$I'_5$ coordinate
system. It is also interesting and significant to apply this
approach to multi-qubit systems, which we shall investigate
subsequently.

 {\bf ACKNOWLEDGMENTS} This work is supported from NSF of China (Grant
No. 10605013) and Program for New Century Excellent Talents in
University.

\end{document}